\documentclass[%
preprint,
superscriptaddress
]{revtex4-1}

\usepackage{graphicx}
\usepackage{dcolumn}
\usepackage{bm}

\usepackage{amsmath}
\usepackage{amssymb}
\usepackage{multirow}
\usepackage{soul}
\usepackage{cancel} 
\usepackage{mdframed}
\usepackage{color}
\setcitestyle{super} 

\begin{document}

\title{Topological nature of bound states in the radiation continuum}
\author{Bo Zhen}
\email{E-mail: bozhen@mit.edu}
\affiliation{Department of Physics, Massachusetts Institute of Technology, 77 Massachusetts Avenue, Cambridge, Massachusetts 02139, USA}

\author{Chia Wei Hsu}
\thanks{These authors contributed equally to this work.}
\affiliation{Department of Physics, Massachusetts Institute of Technology, 77 Massachusetts Avenue, Cambridge, Massachusetts 02139, USA}
\affiliation{Department of Physics, Harvard University, 17 Oxford Street, Cambridge, Massachusetts 02138, USA}

\author{Ling Lu}
\affiliation{Department of Physics, Massachusetts Institute of Technology, 77 Massachusetts Avenue, Cambridge, Massachusetts 02139, USA}

\author{A. Douglas Stone}
\affiliation{Department of Applied Physics, Post Office Box 208284, Yale
University, New Haven, CT 06520, USA}

\author{Marin Solja\v{c}i\'{c}}
\affiliation{Department of Physics, Massachusetts Institute of Technology, 77 Massachusetts Avenue, Cambridge, Massachusetts 02139, USA}

\date{\today}


\maketitle

{\bf Bound states in the continuum (BICs) are unusual solutions of wave equations describing light or matter: they are discrete and spatially bounded, but exist at the same energy as a continuum of states which propagate to infinity.  
Until recently, BICs were constructed through fine-tuning parameters in the wave equation~\cite{1929_Neumann_PZ, 1985_Friedrich_PRA, 2008_Bulgakov_PRB, 2013_Corrielli_PRL, 2013_Weimann_PRL} or exploiting the separability of the wave equation due to symmetry~\cite{1994_Evans, 2011_Plotnik_PRL, 2012_Lee_PRL}. More recently, BICs that that are both robust and not symmetry-protected (``accidental'') have been predicted~\cite{2005_Porter_WM, 2008_Marinica_PRL, 2009_Liu_OE, 2013_Hsu_LSA, 2013_Hsu_Nature} and experimentally realized~\cite{2013_Hsu_Nature} in periodic structures; the simplest such system is a periodic dielectric slab~\cite{2013_Hsu_Nature}, which also has symmetry-protected BICs.
Here we show that both types of BICs in such systems are vortex centers in the polarization direction of far-field radiation. The robustness of these BICs is due to the existence of conserved and quantized topological charges, defined by the number of times the polarization vectors wind around the vortex centers. 
Such charges can only be generated or annihilated by making large changes in the system parameters, and then only according to strict rules, which we derive and test numerically. Our results imply that laser emission based on such states will generate vector beams~\cite{2009_Zhan_AOP}}.  

These BICs exist in photonics crystal slabs, which are dielectric slabs with a periodic modulation of refractive index at the wavelength scale~\cite{JJ_book} surrounded by air or a low-index medium. Photonic crystal slabs are used in many applications including surface-emitting lasers~\cite{2014_Hirose_NatPhoton}, photovoltaics~\cite{2009_Ko_NL},
LEDs~\cite{2009_Wierer_NatPhoton}, and biosensing~\cite{2007_Ganesh_NatNano, 2011_Yanik_PNAS}.
The periodic modulation alters the dispersion relation of light in the slabs and gives rise to photonic bands, analogous to electronic band structures in solids. In general light can escape from the surface of the slab and propagate to the far-field, but a portion of the photonic band states (those ``below the light line") are perfectly confined to the slab by the generalized form of total internal reflection~\cite{JJ_book}.
In contrast, modes above the light line appear generically as resonances with finite lifetimes due to their coupling to the continuum of extended modes~\cite{2002_Fan_PRB}. It has been known for some time that bound states with infinite lifetimes also exist {\it above} the light line at isolated high-symmetry wavevectors. This type of BIC arises from the symmetry mismatch between their mode profiles inside the photonic crystal and those of the external propagating modes (radiation continuum)~\cite{2000_Pacradouni_PRB,2001_Ochiai_PRB,2002_Fan_PRB,2012_Lee_PRL}. Recently, a new type of BIC, unprotected by symmetry, has been found (both theoretically and experimentally) to exist at arbitrary wavevectors in bands above the light line in photonic crystal slabs~\cite{2013_Hsu_Nature}.
These BICs occur ``accidentally'', when the relevant couplings to the continuum (see below) all vanish simultaneously. However their existence does not require fine-tuning of system parameters; small changes in parameters simply shift the position of these special points along the band diagram. An intuitive understanding of why such BICs exist and are robust was previously lacking. Recently, an explanation based on accidental triangular symmetry of the radiating fields was proposed~\cite{2014_Yang_PRL} but does not explain the robustness of these BICs and their occurrences in TE-like bands.


We now show that both types of BICs in photonic crystal slabs are vortex centers in the polarization direction of the far-field radiation of the slabs. 
Using the Bloch theorem for photonic crystals~\cite{JJ_book}, we write the electric field of a resonance as ${\bf E}_{\bf k}({\boldsymbol \rho},z)=e^{i {\bf k}\cdot {\boldsymbol \rho}} {\bf u}_{\bf k}({\boldsymbol \rho},z)$, where ${\bf k}=k_x \hat{x} + k_y \hat{y}$ is the two-dimensional wave vector, ${\boldsymbol \rho} = x \hat{x} + y\hat{y}$ is the in-plane coordinate, ${\bf u}_{\bf k}$ is a periodic function in ${\boldsymbol \rho}$, and $z$ is the normal direction to the slab.
While the fields inside the slab are periodically modulated, outside the slab each state consists of propagating plane waves and/or evanescent waves that decay exponentially away from the surface.
For states above the light line (resonances), and wavelengths below the diffraction limit, the only non-zero propagating-wave amplitudes are the zero-order (constant in-plane) Fourier coefficients of ${\bf u}_{\bf k}$, given by ${\bf c}({\bf k}) = c_{x}({\bf k})\hat{x}+c_{y}({\bf k})\hat{y}$ (Fig.~1a). Here, $c_x({\bf k}) = \hat{x} \cdot \langle {\bf u}_{\bf k} \rangle$, $c_y({\bf k}) = \hat{y} \cdot \langle {\bf u}_{\bf k} \rangle$, and the brackets denote spatial average over one unit cell on any horizontal plane outside the slab. Note that ${\bf c}({\bf k})$ is the projection of $\langle {\bf u}_{\bf k} \rangle$ onto the $xy$ plane; it points in the polarization direction of the resonance in the far field, so we refer to ${\bf c}({\bf k})$ as the ``polarization vector''. 

A resonance turns into a BIC when the outgoing power is zero, which happens if and only if $c_{x}=c_{y}=0$.
In general, $c_{x}$ and $c_{y}$ are both complex functions of ${\bf k}$, and varying the wave vector components $(k_x,k_y)$ is not sufficient to guarantee a solution where  $c_{x}=c_{y}=0$. However, when the system is invariant under the operation $C_2^{z} T$, implying that $\epsilon(x,y,z)=\epsilon^{*}(-x,-y,z)$, we show that $c_x$ and $c_y$ can be chosen to be real numbers simultaneously; in other words, the far field is linearly polarized (see Supplementary Information, here $C_{2}^{z}$ is $180^{\circ}$ rotation operator around $z$ axis, and $T$ is the time reversal operator). When the system also has up-down mirror symmetry ($\sigma_z$), the outgoing waves on one side of the slab determine those on the other;
for such systems, BICs are stable because they correspond to the intersections between the nodal line of $c_x$ and the nodal line of $c_y$ in the $k_x$-$k_y$ plane. Such a nodal intersection naturally causes
a vortex in the polarization vector field centered on the BIC, as illustrated in Fig.~1b, for the simplest case.
Along the nodal line of $c_x$ (or $c_y$), the direction of ${\bf c}({\bf k})$ is along the $y$ axis (or $x$ axis), as illustrated in Fig.~1b. As one encircles the nodal intersection (BIC) in the $k_x$-$k_y$ plane each component
of the polarization vector flips sign as its nodal line is crossed so as to create a net circulation of $\pm 2\pi$ in the polarization field. At the nodal intersection the polarization direction becomes undefined, since at the BIC there is zero emission into the far-field. Conversely one could say that BICs cannot radiate because there is no way to assign a far-field polarization that is consistent with neighbouring ${\bf k}$ points. Thus robust BICs are only possible when there is vorticity in the polarization field.  

Vortices are characterized by their topological charges. Here, the topological charge ($q$) carried by a BIC is defined as:
\begin{equation}
\label{eq:q-main}
q= \frac{1}{2\pi}\oint_C \, d {\bf k} \cdot {\bf \nabla}_{\bf k} \phi ({\bf k}), \quad q \in \mathbb{Z}
\end{equation}
which describes how many times the polarization vector winds around the BIC.
Here, $\phi ({\bf k})=\arg[c_{x}({\bf k})+ic_{y}({\bf k})]$ is the angle of the polarization vector,
and $C$ is a closed simple path in $k$ space that goes around the BIC in the counterclockwise direction.
The fields ${\bf u}_{\bf k}$ are chosen to be smooth functions of ${\bf k}$, so $\phi({\bf k})$ is differentiable in ${\bf k}$ along the path.
The polarization vector has to come back to itself after the closed loop, so the overall angle change must be an integer multiple of $2\pi$, and $q$ must be an integer.
Fig.~1c shows examples of how the polarization vector winds around a BIC with charge $q=+1$ and also around a BIC with charge $q=-1$ along a loop $C$ marked by 1$\rightarrow$2$\rightarrow$3$\rightarrow$4$\rightarrow$1.
Similar definitions of winding numbers as in Eq. 1 can be found in describing topological defects~\cite{1979_Mermin_RMP} of continuous two-dimensional spins, dislocations in crystals, and quantized vortices in helium II~\cite{1991_VortexBook}. This formalism describing polarization vortices is also closely related to Berry phases in describing adiabatic changes of polarization of light~\cite{1987_Berry_JMO} and Dirac cones in gaphenes~\cite{manes2007existence}.



The far-field pattern at a definite ${\bf k}$ point by itself does not reflect the vorticity of polarization around a BIC, but laser emission centered on such a BIC will.
Laser emission always has a finite width in $k$-space and this wave-packet will be centered on the BIC; hence it will consist of a superposition of plane waves from the neighborhood of the BIC, leading naturally to a spatial twist in the polarization for the outgoing beam.  Such beams have been studied previously, and are known as vector beams~\cite{2009_Zhan_AOP}, although their connection with BICs does not appear to have been realized. The number of twists in the polarization direction is known as the order number of the vector beam, and we now see that it is given by the topological charge carried by the BIC. 
%
Note that these vector beams are different from optical vortices, which usually have a fixed polarization direction.

In the example of Fig.~2, we show the topological charges of BICs for a structure that has been experimentally realized in ref.~\citenum{2013_Hsu_Nature}.
In this example, there are five BICs on the lowest-frequency TM-like band (Fig.~2a).  We obtain polarization vectors ${\bf c}({\bf k})$ from finite-difference time-domain (FDTD) calculations, which reveal five vortices with topological charges of $\pm 1$ at these five $k$ points (Fig.~2b).
As discussed above, the BICs and their topological charges can also be identified from the nodal-line crossings and the gray-scale colors of $c_x$ and $c_y$ (Fig.~2c).



The winding number of polarization vector along a closed path is given by the sum of the topological charges carried by all BICs enclosed within this path~\cite{1979_Mermin_RMP}. When system parameters vary continuously, the winding number defined on this path remains invariant, unless there are BICs crossing the boundary. Therefore, topological charge is  a conserved quantity. This conservation rule leads to consequences/restrictions on behaviors of the BICs (Supplementary Information). For example, as long as the system retains $C_2^{z} T$ and $\sigma_z$ symmetries, a BIC can only be destroyed through annihilation with another BIC of the exact opposite charge, or through bringing it outside of the continuum (below the light line). 



The conservation of topological charges allows us to predict and understand the behaviors of BICs when the parameters of the system are varied over a wide range, as we now illustrate. First, consider the lowest-frequency TM-like mode (TM$_{1}$ band) of a 1D-periodic structure in air shown in Fig.~3a. This grating consists of a periodic array of dielectric bars with periodicity of $a$, width $w=0.45a$, and refractive index $n=1.45$. Its calculated band structure is shown in Fig.~3b.  When the thickness of the grating is $h=1.50a$, there are two BICs on the $k_{x}$ axis, as indicated by the radiative quality factor of the resonances (Fig.~3c). The polarization vector ${\bf c}({\bf k})$, also shown in Fig.~3c, characterizes both BICs as carrying charges $q=+1$. When the grating thickness is decreased to $h/a=1.43$ (all other parameters fixed), the two BICs move towards the center of the Brillouin zone, meet at the ${\bf \Gamma}$ point, and deflect onto the $k_{y}$ axis (Fig.~3d). This is inevitable due to the conservation of the topological charges: annihilation cannot happen between two BICs of the same charge.

Annihilation of BICs is only possible when charges of opposite signs are present. This can be seen in the lowest-frequency TE-like band of the same structure (Fig.~3e,f).  When $h/a=1.04$, there are two off-${\bf \Gamma}$ BICs with charge $-1$ and a BIC with charge $+1$ at the ${\bf \Gamma}$ point (Fig.~3e). As $h/a$ decreases, the two $-1$ charges move to the center and eventually annihilate with the $+1$ charge, leaving only one BIC with charge $q=-1$ (Fig.~3f).


Generation of BICs is also restricted by charge conservation, and can be understood as the reverse process of annihilation. We provide an example by considering the lowest-frequency TE-like mode in a photonic crystal slab of $n=3.6$ with a square lattice of cylindrical air holes of diameter $d=0.5a$ (Fig.~4a). As the slab thickness increases, BICs are generated at the ${\bf \Gamma}$ point.  Each time, four pairs of BICs with exact opposite charges are generated, consistent with charge conservation and $C_{4v}$ symmetry of the structure. With further increase of the slab thickness, the eight BICs move outward along high-symmetry lines and eventually go outside of the continuum (fall below the light line).

Although the examples discussed so far only show topological charges of $\pm1$, other values of charges can be found in higher-frequency bands of the PC or in structures with higher rotational symmetry. For example, Fig.~S2 in Supplementary Information shows a stable BIC of charge $-2$ at the ${\bf \Gamma}$ point arising from the double degeneracy of nodal lines caused by the $C_{6v}$ symmetry of the system.

The symmetries of the system also restrict the possible values of topological charges, since the nodal curves must respect the point symmetry. BICs at ${\bf k}$ points related by in-plane point-group symmetries of the system (mirror reflections and rotations) have the same topological charges (Supplementary Information).
Using this fact, one can calculate all possible topological charges of BICs at high symmetry ${\bf k}$ points. For a system with $C_{n}$ symmetry, the possible topological charges at the $\Gamma$ point on a singly-degenerate band are given in Table S1 (Supplementary information). This is consistent with all examples in this paper. This table can be used to predict the charges in other systems of interest and to design high order vector beams.  
%


{\bf{Conclusions}}

We have demonstrated that BICs in photonic crystal slabs are associated with vortices in the polarization field and explained their robustness in terms of 
conserved topological charges. We derive the symmetries that constrain these charges and explain their generation, evolution and annihilation. We conjecture that all robust BICs~\cite{1994_Evans, 2011_Plotnik_PRL, 2012_Lee_PRL, 2005_Porter_WM, 2008_Marinica_PRL, 2009_Liu_OE, 2013_Hsu_LSA, 2013_Hsu_Nature}  will correspond to vortices in an appropriate parameter space. Our finding connects electromagnetic BICs to a wide range of physical phenomena including Berry phases around Dirac points~\cite{manes2007existence}, topological defects~\cite{1979_Mermin_RMP}, and general vortex physics~\cite{1991_VortexBook}.
Optical BICs in photonic crystals have a wealth of applications. Lasing action can naturally occur at BIC states where the quality factor diverges. The angular~(wavevector) tunablity of the BICs makes them great candidates for on-chip beam-steering~\cite{2010_Kurosaka_NaturePhoton}.
Furthermore, photonic crystal lasers through BICs are naturally vector beams~\cite{2011_Iwahashi_OE,2012_Kitamura_OL}, which are important for particle accelerations, optical trapping and stimulated emission depletion microscopy. 


{\bf Acknowledgments} 

The authors thank Chong Wang, Scott Skirlo, David Liu, Fan Wang, Nicholas Rivera, Dr. Ido Kaminer, Dr. Homer Reid, Prof. Xiaogang Wen, Prof. Steven G. Johnson, and Prof. John D. Joannopoulos for helpful discussions. This work was partly supported by the Army Research Office through the Institute for Soldier Nanotechnologies under contract no. W911NF-07-D0004. B.Z., L.L., and
M.S. were partly supported by S3TEC, an Energy Frontier Research Center funded by
the US Department of Energy under grant no. DE-SC0001299. L.L. was supported in part by the Materials Research Science and Engineering Center of the National Science Foundation (award no. DMR-0819762). A.D.S. was partly supported by NSF grant DMR-1307632.

{\bf Author contributions} 

All authors discussed the results and made critical contributions to the work.

{\bf Additional information} 

Supplementary information is available in the online version of the paper. Reprints and permissions information is available online at www.nature.com/reprints. Correspondence and requests for materials should be addressed to B.Z.

{\bf Competing financial interests}

The authors declare no competing financial interests.

\bibliographystyle{naturemag}
\bibliography{mybib}

\clearpage

{\bf Supplementary information}

\renewcommand{\theequation}{S.\arabic{equation}}
\setcounter{equation}{0}

\subsection{Symmetry requirements for stable BICs}

Here, we give the proof that stable BICs at arbitrary $k$ points can be found when the system is invariant under $C_{2}^{z}T$ and $\sigma_{z}$ operators, and that stable BICs at $C_{2}^{z}$-invariant $k$ points can be found when the system has $C_{2}^{z}$ symmetry. Here, $C_{2}^{z}$ means $180^{\circ}$ rotation around $z$ axis, and $T$ means the time reversal operator. The schematics of the symmetry requirement is summarized in Fig. S1.



\textbf{In region \textit{I}}, systems are invariant under the symmetry operator $C_{2}^{z}T$, namely $\epsilon^{\star}(x,y,z)=\epsilon(-x,-y,z)$. 
Let ${\bf u}_{\bf k}$ be an eigenfunction of the master operator~\cite{JJ_book} $\Theta_{\bf k} = \frac{1}{\epsilon} (\nabla+i{\bf k}) \times (\nabla+i{\bf k}) \times$, and recall that ${\bf k}$ here only has $x$ and $y$ components since we are considering a slab structure that does not have translational symmetry in $z$. A short derivation shows that at any ${\bf k}$ point, ${\bf u}_{\bf k}({\bf r})$ and $C_2^z {\bf u}_{\bf k}^*(C_2^z{\bf r})$ are both eigenfunctions of $\Theta_{\bf k}({\bf r})$ with the same eigenvalue, so they must differ at most by a phase factor,
\begin{eqnarray}
\label{eq:C2T}
{\bf u}_{\bf k}({\bf r}) &=& e^{i\theta_{\bf k}}C_2^z {\bf u}_{\bf k}^*(C_2^z{\bf r}) \nonumber \\ 
 &=&  e^{i\theta_{\bf k}}(-{\bf u}_{\bf k}^{x*},-{\bf u}_{\bf k}^{y*},{\bf u}_{\bf k}^{z*}) \rvert_{(-x,-y,z)}
\end{eqnarray}
Here $\theta_{\bf k}$ is an arbitrary phase factor. 
Meanwhile, we are free to multiply ${\bf u}_{\bf k}$ with any phase factor, and it remains a valid eigenfunction. For our purpose here, we explicitly choose the phase factor of ${\bf u}_{\bf k}$ such that $e^{i\theta_{\bf k}}=-1$ for all ${\bf k}$. With this choice, we can average over $x$ and $y$ to get ${\bf c}({\bf k}) = {\bf c}^*({\bf k})$ for all ${\bf k}$. That is, the polarization vector ${\bf c}({\bf k})$ is purely real. 

Using the fact that systems in region {\it I} also have the up-down mirror symmetry $\sigma_{z}$, namely $\epsilon(x,y,z)=\epsilon(x,y,-z)$, we can link the radiation loss above and below the photonic crystal slab denoted by ${\bf c}^{\uparrow}$ and ${\bf c}^{\downarrow}$.  
At any ${\bf k}$ point, ${\bf u}_{\bf k}({\bf r})$ and $\sigma_z {\bf u}_{\bf k}(\sigma_z{\bf r})$ are both eigenfunctions of $\Theta_{\bf k}({\bf r})$ with the same eigenvalue, so
\begin{eqnarray}
\label{eq:sigma_z}
{\bf u}_{\bf k}({\bf r}) &=& e^{i\theta_{\bf k}} \sigma_z {\bf u}_{\bf k}(\sigma_z{\bf r}) \nonumber \\ 
 &=&  e^{i\theta_{\bf k}}({\bf u}_{\bf k}^{x},{\bf u}_{\bf k}^{y},-{\bf u}_{\bf k}^{z}) \rvert_{(x,y,-z)}
\end{eqnarray}
with $\theta_{\bf k}$ being an arbitrary phase factor (not to be confused with the one in Eq.~\eqref{eq:C2T}).
Since $\sigma_z^2 = 1$, we can apply Eq.~\eqref{eq:sigma_z} twice to show that $e^{i\theta_{\bf k}} = \pm 1$. Averaging over $x$ and $y$, we see that ${\bf c}^{\uparrow} = \pm {\bf c}^{\downarrow}$.

After using these two symmetries, the number of independent real variables in all radiation coefficients $c^{\uparrow,\downarrow}_{x,y}$ has been reduced from 8 to 2. Given that the number of independent tuning parameters is also 2: ($k_{x},k_{y}$), we are able to get stable BICs.
Note that the combination of $C_{2}^{z}T$ and $\sigma_{z}$ is just one sufficient condition for stable BICs in photonic crystal slabs. There might be other different choices of symmetries. For example, $PT$ and $\sigma_{z}$ is equivalent to $C_{2}^{z}T$ and $\sigma_{z}$, where $P$ is the inversion operator. Also, the requirement of $\sigma_{z}$ is not necessary when there is leakage to one direction only (such as BICs on the surface of a photonic bandgap structure~\cite{2013_Hsu_LSA}).

\textbf{In region \textit{II}}, stable BICs at $C_{2}^{z}$-invariant $k$ points can be found. Systems in this region have $C_{2}^{z}$ symmetry, namely  $\epsilon(x,y,z)=\epsilon(-x,-y,z)$. $k$ points are $C_{2}^{z}$-invariant when $-\bf{k}=\bf{k}+\bf{G}$, with $\bf{G}$ being a reciprocal lattice vector.  A short derivation shows that at any $k$ point, ${\bf u}_{\bf k}({\bf r})$ and $C_2^z {\bf u}_{-{\bf k}}(C_2^z{\bf r})$ are both eigenfunctions of $\Theta_{\bf k}({\bf r})$ with the same eigenvalue, so
\begin{equation}
{\bf u}_{{\bf k}}({\bf r}) = e^{i\theta_{\bf k}} C_2^z {\bf u}_{-{\bf k}}(C_2^z {\bf r}),
\label{eq:C2z}
\end{equation}
with $\theta_{\bf k}$ being an arbitrary phase factor (not to be confused with the two phase factors above).
At these high-symmetry ${\bf k}$ points, using Bloch theorem we know: ${\bf u}_{-{\bf k}} = {\bf u}_{{\bf k}+{\bf G}} = {\bf u}_{{\bf k}}$, so we can apply Eq.~\eqref{eq:C2z} twice to get $e^{i\theta_{\bf k}} = \pm 1$. When this factor is $+1$, we can average over $x$ and $y$ to see that ${\bf c}({\bf k})=0$, corresponding to a BIC at this $C_{2}^{z}$-invariant $k$ point. 

\textbf{In region \textit{III}}, both kinds of BICs can be found, where $C_{2}^{z}$, $T$ and $\sigma_{z}$ are all present. All our numerical examples are within this region to make it easier to understand the relation and interaction between different types of BICs. 

\subsection{Consequences of topological charge conservation} 

Since topological charge is a conserved quantity, there are a few consequences and restriction on the evolution of BICs. First, BICs are stable as long as the system retains required symmetries;  however, perturbations that break these two required symmetries eliminate the existence of BICs. When $C_{2}^{z}T$ symmetry is broken, the coefficients ($c_{x}$ and $c_{y}$) require complex components, meaning the radiation becomes elliptically polarized instead of linearly polarized. When $\sigma_{z}$ symmetry is broken, the coefficients $c^{\uparrow,\downarrow}_{x,y}$ are still real numbers, but radiation towards the top and towards the bottom become separate degrees of freedom and so they do not vanish simultaneously in general. Second, when BICs collide into each other in the moment space, the sum of all topological charges they carry remains the same before and after the collision.

\subsection{Example of charge -2}

We consider the lowest-frequency TE-like mode of a photonic crystal slab with a hexagonal lattice of cylindrical air holes (shown in Fig. S2a). The refractive index of the slab is $n=1.5$; the air-hole diameter is $0.5a$; and the thickness of the slab is 0.5a, where $a$ is the lattice constant.  This system has $C_{6}^{z}$ symmetry. Normalized lifetime plot indicates a BIC at the center of the Brillouin zone shown in Fig. S2b. The polarization vector field characterizes the BIC carrying charge $-2$ shown in Fig. S2c. Charge $-2$ can also be understood from the double degeneracy of both nodal lines of $c_{x}$ (green) and $c_{y}$ (red), shown in the inset of Fig. S2c. All four nodal lines are pinned at $\Gamma$ point stabilized by the $C_{6}$ symmetry. 

\subsection{BICs related by point group symmetries have the same topological charges}

Here, we prove that when the structure has a certain in-plane point group symmetry $\mathcal{R}$ (namely, $\epsilon({\bf r})=\epsilon(\mathcal{R}{\bf r})$; $\mathcal{R}$ can be a combination of rotation and reflection on the $x$-$y$ plane) and when the band has no degeneracy, a BIC at ${\bf k}$ indicates there is another BIC at $\mathcal{R}{\bf k}$ with the same topological charge.
The assumption here is that the eigenfunctions ${\bf u}_{\bf k}$ at different $k$ points already have their phases chosen to ensure the reality of ${\bf c}({\bf k})$, and the signs of ${\bf u}_{\bf k}$ at different $k$ points have been chosen such that ${\bf u}_{\bf k}$ is continuous with respect to ${\bf k}$ (so that a small change in ${\bf k}$ leads to a small change in ${\bf u}_{\bf k}$).

We start by relating the eigenfunction at ${\bf k}$ and the eigenfunction at $\mathcal{R}\bf{k}$.
Let ${\bf u}_{\bf k}$ be an eigenfunction of operator $\Theta_{\bf k}$. Since the system is invariant under transformation $\mathcal{R}$, we know $\hat{O}_{\mathcal{R}}{\bf u}_{\bf k}$ is an eigenfunction of $\Theta_{\mathcal{R}{\bf k}}$, so in the absence of degeneracy, we can write $\hat{O}_{\mathcal{R}}{\bf u}_{\bf k} = \alpha_{\bf k} {\bf u}_{\mathcal{R}{\bf k}}$, where $\alpha_{\bf k}$ is some number.
The number $\alpha_{\bf k}$ must have unit magnitude (due to the normalization of ${\bf u}_{\bf k}$ and ${\bf u}_{\mathcal{R}{\bf k}}$) and must be real-valued (because ${\bf c}({\bf k})$ is real-valued), so it can only take on discrete values of $\pm 1$.
Also, $\alpha_{\bf k}$ must be a continuous function of ${\bf k}$ since ${\bf u}_{\bf k}$ is continuous with respect to ${\bf k}$.
Since $\alpha_{\bf k}$ is both discrete-valued and continuous, it must be a constant. 
Then, we may denote this constant with its value at the $\Gamma$ point, as $\alpha_{\bf k} = \alpha_{\Gamma}$. 
 Note that ${\mathcal{R}} {\bf \Gamma}={\bf \Gamma}$, so we can determine coefficient $\alpha_{\bf \Gamma}$ using the mode profile: $\hat{O}_{\mathcal{R}}{\bf u}_{\bf \Gamma} = \alpha_{\bf \Gamma} {\bf u}_{\bf \Gamma}$. 
In conclusion, we have ${\bf u}_{\mathcal{R}{\bf k}} = \alpha_{\Gamma} \hat{O}_{\mathcal{R}}{\bf u}_{\bf k}$.

Now we consider how the angle $\phi(\mathcal{R}{\bf{k}})$ is related to $\phi(\bf{k})$. The vector field ${\bf u}_{\bf k}$ transforms under the rotation operator as $( \hat{O}_{\mathcal{R}} {\bf u}_{\bf k} ) ({\bf r})=\mathcal{R}{\bf{u}}_{\bf{k}}(\mathcal{R}^{-1}\bf r)$, so averaging over $x$ and $y$ we get $\langle \hat{O}_{\mathcal{R}} {\bf u}_{\bf k} \rangle = \mathcal{R} \langle {\bf u}_{\bf k} \rangle$.
Let $P$ be the operator that projects a 3D vector onto the $x$-$y$ plane, namely $P{\bf r} = {\bf r} - ({\bf r} \cdot \hat{z})\hat{z}$; it commutes with $\mathcal{R}$, since it does not alter the $x$ or $y$ component. Then ${\bf c}({\bf k}) = P \langle {\bf u}_{\bf k} \rangle$, and
\begin{equation}
\label{eq:s4}
{\bf c}(\mathcal{R}{\bf{k}})
= P \langle  {\bf u}_{\mathcal{R} {\bf k}} \rangle
= P \langle  \alpha_{\Gamma} \hat{O}_{\mathcal{R}}{\bf u}_{\bf k}  \rangle
= \alpha_{\Gamma}  \mathcal{R}  P \langle{\bf{u}}_{\bf{k}} \rangle 
= \alpha_{\Gamma}  \mathcal{R} {\bf c}({\bf{k}}).
\end{equation}
So, the polarization vector at the the transformed $k$ point is simply the original polarization vector transformed and times $\pm 1$. So, the angle of the polarization vector only changes by a constant in the case of proper rotations (where $\det{\mathcal{R}}=1$); in the case of improper rotations (where $\det{\mathcal{R}}=-1$), it also changes sign. So, in general, we can write
\begin{equation}
\phi(\mathcal{R}{\bf k}) = (\det{\mathcal{R}}) \phi({\bf k}) + c 
\end{equation}
with $c$ being a constant depending on $\mathcal{R}$ and $\alpha_{\Gamma}$. It follows that $\nabla_{\mathcal{R}{\bf k}} \phi(\mathcal{R}{\bf k})  = (\det{\mathcal{R}}) \mathcal{R} \nabla_{\bf k} \phi({\bf k}) $, so the topological charge at $\mathcal{R}{\bf k}$ is
\begin{align}
\label{eq:q}
q_{\mathcal{R}{\bf k}} &= \frac{1}{2\pi}\oint_{C_{\mathcal{R}{\bf k}}} \, {\bf \nabla}_{{\bf k}''} \phi ({\bf k}'') \cdot d {\bf k}'' \nonumber \\ 
 &= \frac{1}{2\pi}\oint_{\mathcal{R}^{-1}C_{\mathcal{R}{\bf k}}} \, {\bf \nabla}_{\mathcal{R}{\bf k}'} \phi (\mathcal{R} {\bf k}') \cdot \mathcal{R} d {\bf k}' \nonumber \\ 
 &= \frac{1}{2\pi}  (\det{\mathcal{R}}) \oint_{C_{{\bf k}}} \, {\bf \nabla}_{\mathcal{R}{\bf k}'} \phi (\mathcal{R} {\bf k}') \cdot \mathcal{R} d {\bf k}' \nonumber \\ 
&=  \frac{1}{2\pi}  (\det{\mathcal{R}})^{2} \oint_{C_{{\bf k}}} \,  \cancel{\mathcal{R}} {\bf \nabla}_{{\bf k}'} \phi({\bf k}') \cdot \cancel{\mathcal{R}} d {\bf k}' \nonumber \\ 
 &= q_{\bf k},
\end{align}
where $C_{\mathcal{R}{\bf k}}$ is a closed simple path that is centered on $\mathcal{R}{\bf k}$ and loops in counterclockwise direction, $\mathcal{R}^{-1}C_{\mathcal{R}{\bf k}}$ is this loop transformed by $\mathcal{R}^{-1}$ (which centers on ${\bf k}$ in counterclockwise direction if $\mathcal{R}$ is a proper rotation, or in clockwise direction if $\mathcal{R}$ is improper), and $C_{{\bf k}}$ is this transformed loop traversed in counterclockwise direction.

In conclusion, we have proven that if a system has certain point group symmetry $\mathcal{R}$, then the topological charges carried by the BIC at $\bf k$ and at $\mathcal{R}{\bf k}$ on a singly degenerate band have to be the same. This conclusion agrees with all examples in Figs. 2-4.

\subsection{Allowed charges at $\Gamma$ in systems with different symmetries}

Allowed topological charges at high symmetry $k$ points can be determined by the field eigenvalues of the rotational symmetry of a system. For systems with $m$-fold rotational symmetry, we can first determine the relationship between polarization direction at wavevector ${\bf k}$ and at rotated wavevector  $\mathcal{R}{\bf k}$ ($\phi({\bf k})$ and  $\phi(\mathcal{R}{\bf k})$) using Eq.~\eqref{eq:s4}. Since the wavevector gets back to its original point if applying this rotation $m$-times: $\mathcal{R}^{m}{\bf k}={\bf k}$, we can then apply this relationship $m$ times and get how many times the polarization vector rotates around the center of the Brillouin zone. From there, we categoraize all possible charges allowed at ${\bf \Gamma}$ as shown in Table S1. Allowed charges depend on two factors. The first one is which symmetry representation the band belongs to. The second one is the degeneracy of nodal lines at $\Gamma$, because more nodal lines intersecting at the same point usually leads to more oscillations in color and thus higher topological charges. This factor is reflected by the integer number $n$, depending on the number of equivalent ${\bf \Gamma}$ points at this frequency~\cite{2011_Iwahashi_OE}.   
 Note that only singly degenerate bands are considered in this Letter, having no crossing with other bands in the bandstructures, as can be seen in Table S1. Further research directions may include BICs on degenerate bands, as well as the search of BICs with higher-order and potentially fractional topological charges.


\begin{table}[h]
\begin{tabular}{|c|c|c|c|c|}
\hline
Symmetries                  & Representation     & Charges & Allowed $n$             & Allowed charges \\ \hline
\multirow{2}{*}{$C_{2}$} & A     & $\pm1+2n$   & \multirow{2}{*}{$0$}    & $\pm1$              \\ \cline{2-3} \cline{5-5} 
                    & B     & $0+2n$    &                       & 0               \\ \hline
$C_{3}$                  &A    & $1+3n$    & $0,\pm1,...$                  & +1,+4,-2,...    \\ \hline
\multirow{2}{*}{$C_{4}$} & A & $1+4n$    & \multirow{2}{*}{$0,\pm1,...$ } & +1,+5,-3,...    \\ \cline{2-3} \cline{5-5} 
                    & B & $-1+4n$   &                       & -1,-5,+3,...    \\ \hline
\multirow{2}{*}{$C_{6}$} & A & $1+6n$   & \multirow{2}{*}{$0,\pm1,...$ } & +1,+7,-5,...    \\ \cline{2-3} \cline{5-5} 
                    & B & $-2+6n$   &                       &  -2,+4,-8,...   \\ \hline
\end{tabular}
\caption{Allowed stable topological charges at $\Gamma$ for singly degenerate bands. $A(B)$ corresponds to modes of different representations of the symmetry operator~\cite{2005_Sakoda_Book}.  Note that only singly degenerate representations of symmetry operators are included in here.}
\end{table}

\begin{figure*}[ht]
\includegraphics[width=\textwidth]{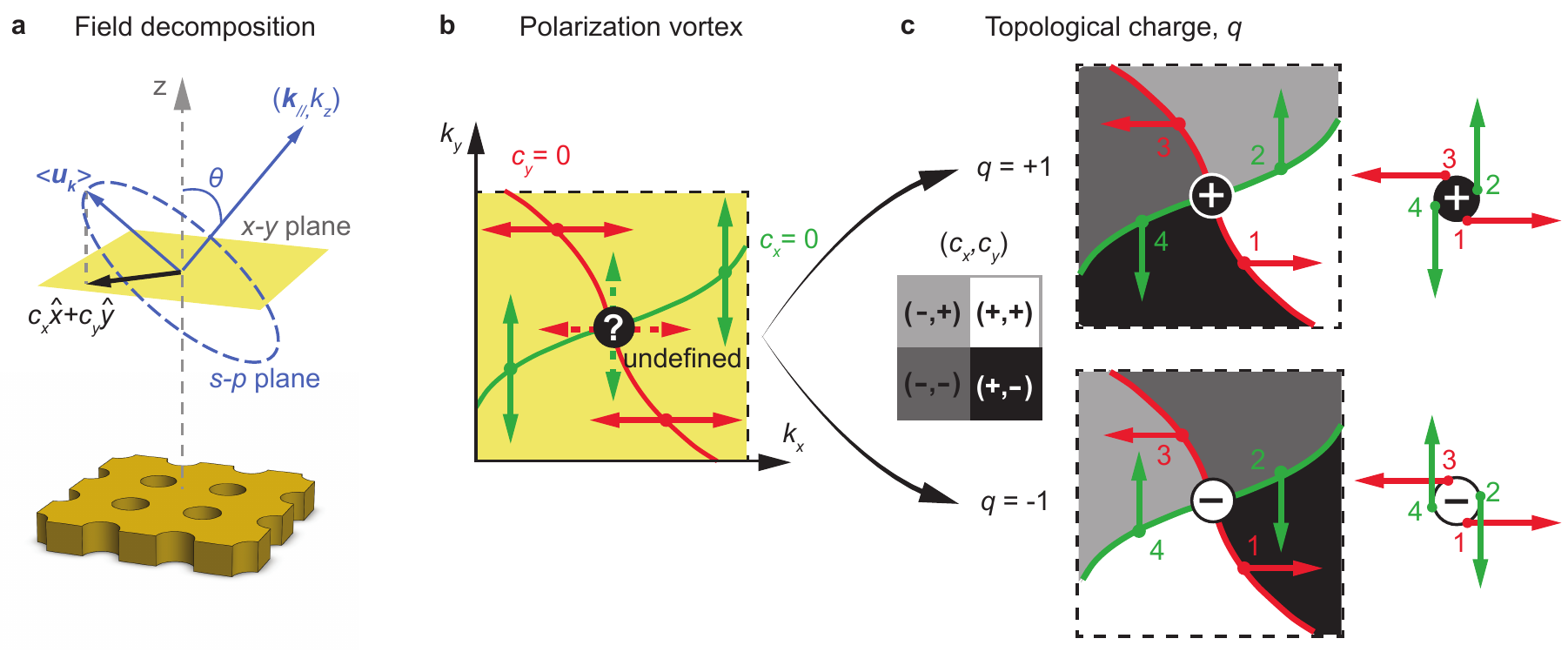}
\caption{{\bf Stable bound states in the continuum (BICs) as vortex centers of polarization vectors.}
{\bf a}, Schematics of radiation field decomposition for resonances of a slab structure. The spatially-averaged Bloch part of the electric field $\langle \textbf{\textit{u}}_{\textbf{\textit{k}}} \rangle$ is projected onto the $x$-$y$ plane as the polarization vector $\textbf{\textit{c}}=(c_{x},c_{y})$. A resonance turns into a BIC if and only if $c_{x}=c_{y}=0$.
{\bf b}, Schematic illustration for the nodal lines of $c_{x}$ (green) and of $c_{y}$ (red) in a region of ${\bf k}$ space near a BIC. The direction of vector $\textbf{\textit{c}}$ (shown in arrows) becomes undefined at the nodal line crossing, where a BIC is found.
{\bf c}, Two possible configurations of the polarization field near a BIC. 
Along a closed loop in $k$-space containing a BIC (loop goes in counterclockwise direction, 1$\rightarrow$2$\rightarrow$3$\rightarrow$4), the polarization vector either rotates by angle $2\pi$ (denoted by topological charge $q=+1$) or rotates by angle $-2\pi$ (denoted by topological charge $q=-1$).
Different regions of the $k$ space are colored in four gray-scale colors according to the signs of $c_{x}$ and $c_{y}$. In this way, a BIC happens where all four gray-scale colors meet, and charge $q=+1$ corresponds to the color changing from white to black along the counterclockwise loop $C$, and charge $q=-1$ corresponds to the color changing from black to white.
}
\label{fig1}
\end{figure*}

\begin{figure*}[hb]
\includegraphics[width=\textwidth]{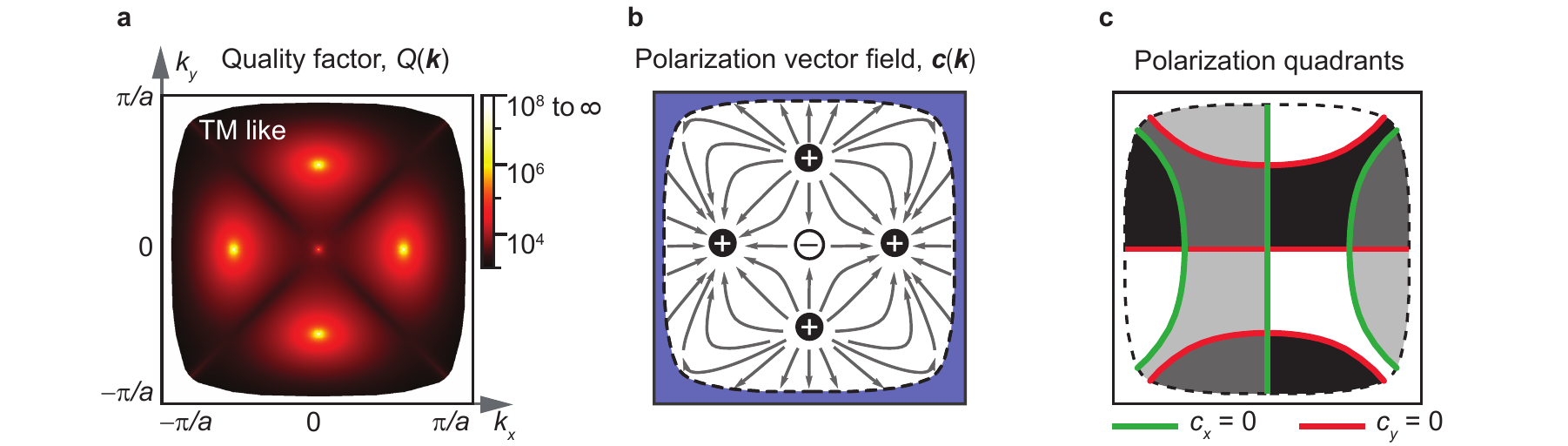}
\caption{{\bf Characterization of BICs using topological charges.}
{\bf a}, Calculated radiative quality factor $Q$ of the TM$_1$ band on a square-lattice photonic crystal slab (as in ref.~\citenum{2013_Hsu_Nature}), plotted in the first Brillouin zone. Five BICs can be seen.
{\bf b}, Directions of the polarization vector field reveal vortices with topological charges of $\pm 1$ at each of the five $k$ points. The area shaded in blue indicates modes below the lightline and thus bounded by total internal reflection. 
{\bf c}, Nodal lines and gray-scale colors of the polarization vector fields (same coloring scheme as in Fig. 1c).}
\label{fig2}
\end{figure*}

\begin{figure*}[ht]
\begin{centering}
\includegraphics[width=0.5\textwidth]{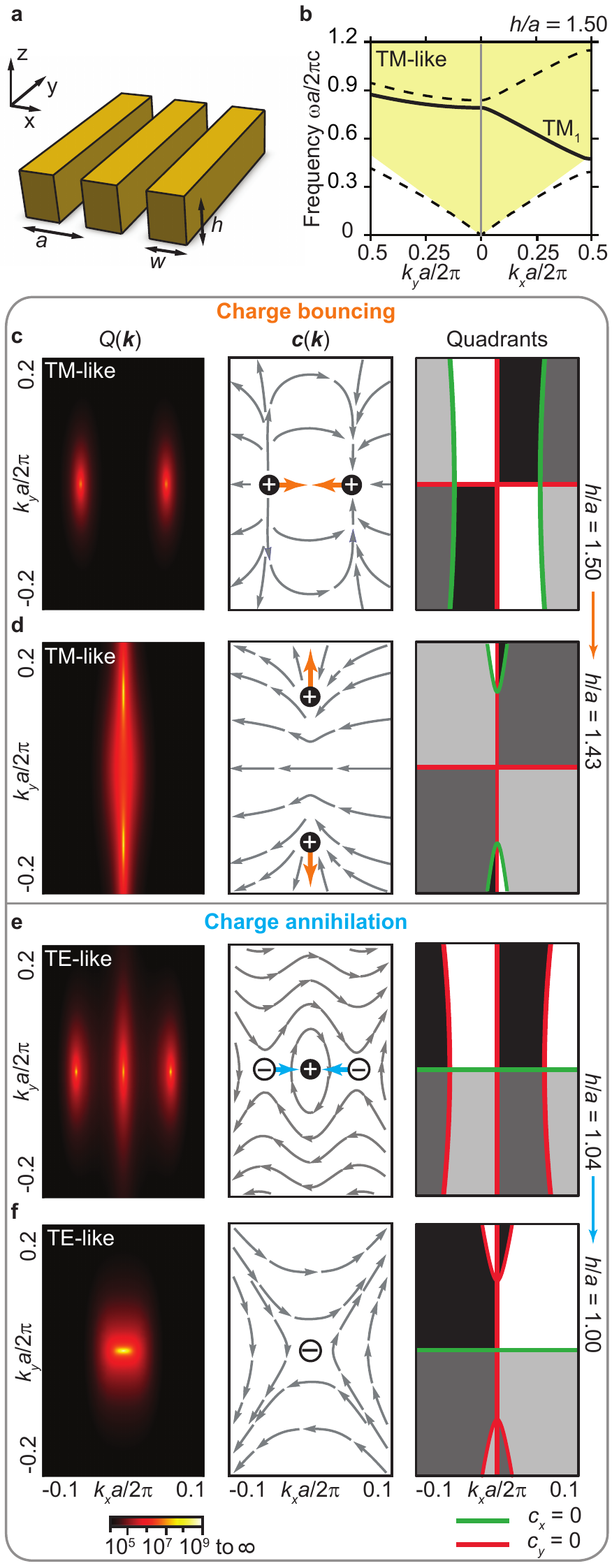}
\label{fig3}
\end{centering}
\end{figure*}
\clearpage
{\baselineskip16pt {FIG. 3. {\bf Evolution of BICs and conservation of topological charges.}}
{\bf a}, Schematic drawing of a photonic crystal slab with one-dimensional periodicity in $x$ and is infinitely long in $y$.
{\bf b}, Calculated TM-like band structure along $k_x$ axis and along $k_y$ axis. The area shaded in yellow indicates the light cone, where there is a continuum of radiation modes in the surrounding medium. 
{\bf c, d}, An example showing topological charges with the same sign bouncing off each other. 
As the slab thickness $h$ decreases, the two BICs with charge $+1$ move along the $k_x$ axis, meet at the origin, and then deflect onto the $k_y$ axis. This can be understood from the conservation of topological charges or from the evolution of nodal lines.
{\bf e, f}, An example of topological charge annihilation happening on the lowest-frequency TE-like band of the same structure with different slab thicknesses.
As the slab thickness $h$ decreases, two BICs with charge $-1$ meet with a BIC with charge $+1$ at the origin. These three BICs annihilate to yield one BIC with charge $-1$, as governed by charge conservation.}

\begin{figure*}[h]
\setcounter{figure}{3}
\includegraphics[width=0.5\textwidth]{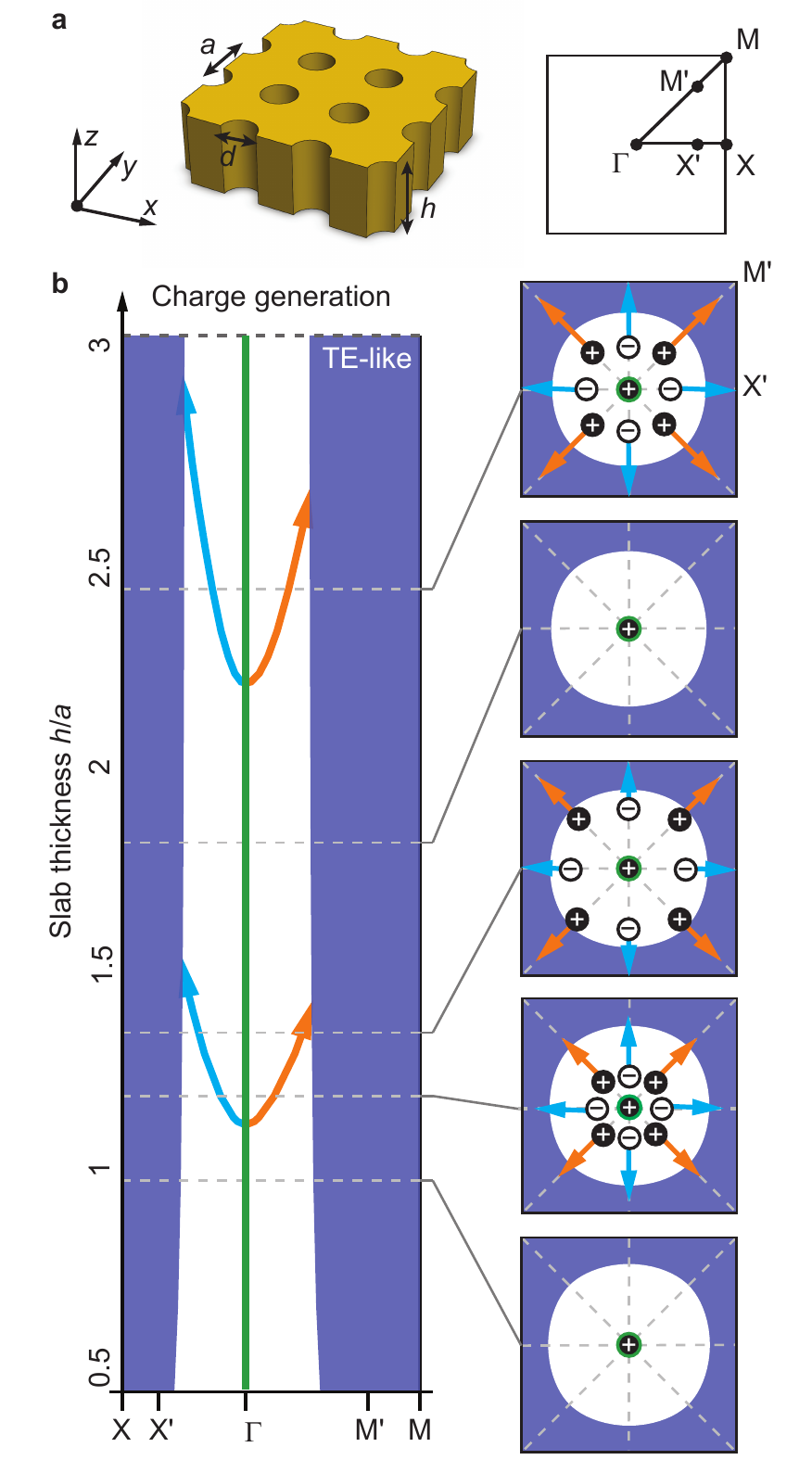}
\caption{{\bf Generation of BICs.}
{\bf a}, Schematic drawing of a photonic crystal slab with two-dimensional periodicity.
{\bf b}, Generation of BICs on the TE$_1$ band when the slab thickness $h$ is increased. Each time, four pairs of BICs with charges $\pm 1$ are generated simultaneously, consistent with the charge conservation and $C_{4v}$ symmetry.
Insets show the locations of BICs in the $k$ space and their corresponding topological charges for $h/a = 1.0, 1.2, 1.35, 1.8,$ and $2.4$. As the slab thickness increases, the BICs move outward and eventually fall below the light line into the area shaded in dark blue.  
}
\label{fig4}
\end{figure*}

\renewcommand{\figurename}{{FIG.}}
\renewcommand{\thefigure}{{ S\arabic{figure}}}
\setcounter{figure}{0}
\begin{figure*}[ht]
\includegraphics[width=0.5\textwidth]{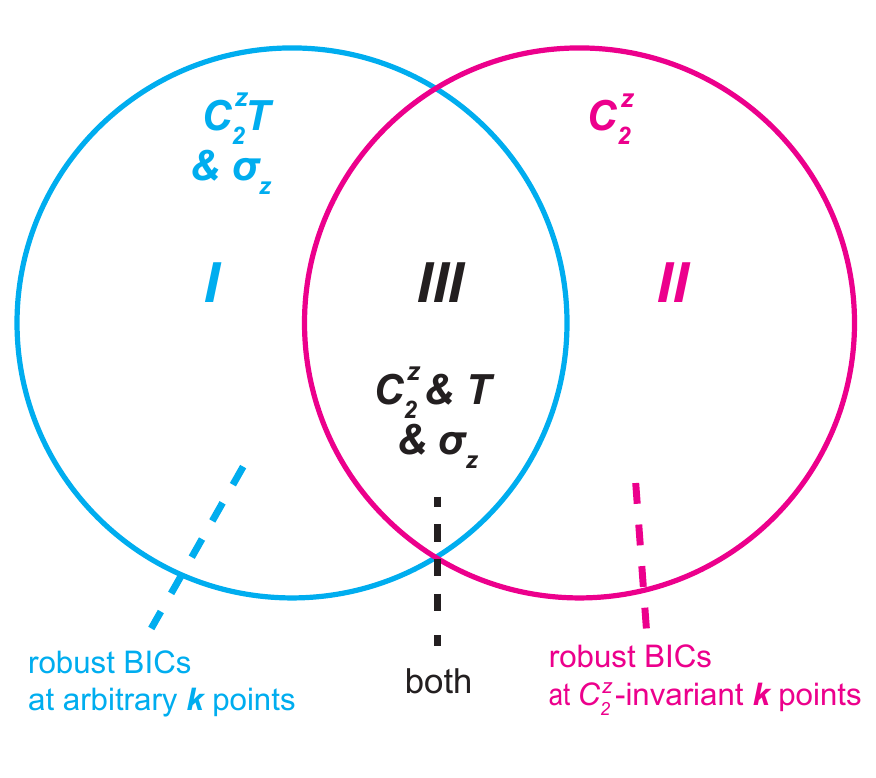}
\caption{{\bf Symmetry requirements for BICs.}
Systems in the blue circle are invariant under operators $C_{2}^{z}T$ and $\sigma_{z}$, where stable BICs at arbitrary wavevectors can be found. In the red circle, where $C_{2}^{z}$ is a symmetry of the system, robust BICs can be found at high-symmetry wavevector points.  Here, high-symmetry wavevectors mean $C_{2}^{z}$-invariant ones, while arbitrary wavectors are not necessarily $C_{2}^{z}$-invariant. In the overlapping area (region {\it III}), both types BICs can be found. All numerical examples in this Letter are within region {\it III}. }
\label{figS1}
\end{figure*}

\begin{figure*}[b]
\includegraphics[width=\textwidth]{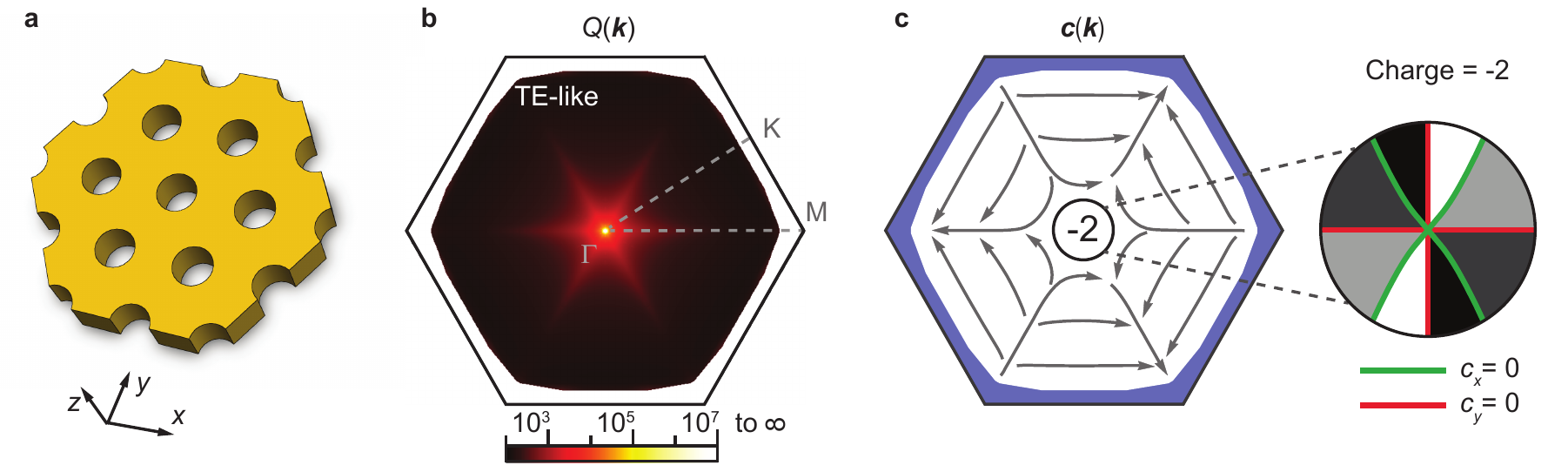}
\caption{{\bf Stable BIC with topological charge -2.}
{\bf a}, Schematic drawing of the photonic crystal slab.
{\bf b}, $Q$ plotted in the first Brillouin zone, showing a BIC at the $\Gamma$ point.
{\bf c}, Polarization vector field characterizes the BIC with a stable topological charge of -2, as can be shown from double degeneracies of both nodal lines. }
\label{figS1}
\end{figure*}


\end{document}